\documentclass[%
 reprint,
 amsmath,amssymb,
 aps,
 pra,
]{revtex4-2}

\usepackage{graphicx}
\usepackage{dcolumn}
\usepackage{bm}
\usepackage{makecell}
\usepackage{lineno}
\usepackage{float}
\usepackage[caption=false]{subfig}

\begin{document}

\preprint{APS/123-QED}

\title{Expansion Dynamics of Cold Non-Neutral Rubidium Plasma}

\author{Michael A. Viray}
\author{Stephanie A. Miller}
\altaffiliation{Present Address: Terumo Heart, Inc., Ann Arbor, MI 48103, USA}
\author{Georg Raithel}%
\affiliation{%
 Department of Physics, University of Michigan, Ann Arbor, Michigan 48109, USA
}%

\date{\today}

\begin{abstract}
We study the expansion of a strongly coupled, non-neutral, and cylindrically arranged ion plasma into the vacuum. The plasma is made from cold rubidium atoms in a magneto-optical trap (MOT) and is formed via ultraviolet photoionization. Higher-density and lower-density plasmas are studied to exhibit different aspects indicative of strong coupling. In a higher-density regime, we report on the formation and persistence of plasma shock fronts, as well as external-field-induced plasma focusing effects. In the lower-density regime, conditions are ideal to observe the development and evolution of nearest-neighbor ion correlations, as well as geometry-induced asymmetries in the pair correlation function. Simulated results from a trajectory model and a fluid model are in good agreement with the measurements.
\end{abstract}

\maketitle


\section{\label{sec:level1}Introduction}

The development of atom trapping and cooling processes \cite{metcalf} has made it possible for researchers to photo-excite cold plasmas in the laboratory \cite{killian99, kulin00, pohl03, pohl05}. This enables models that scale to hard-to-access plasmas that occur, for instance, in astrophysical environments (insides of stars and gas planets) \cite{woolsey}, magnetic-confinement fusion \cite{mckenna, wessels18}, and inertial-confinement fusion \cite{djotyan18, jiang19}. The low temperatures allow experimenters to produce and study strongly coupled plasmas. Two parameters commonly used to quantitatively characterize a plasma are the Debye length, $\lambda_D$, and the Coulomb coupling parameter, $\Gamma$. The Debye length is a measure for a plasma particle's electrostatic shielding within the plasma \cite{bittencourt} and is defined as

\begin{equation}
    \lambda_D = \sqrt{\frac{\varepsilon_0 k_B T}{n q^2}}
    \label{eq:debye}
\end{equation}

Here, $\varepsilon_0$ is the vacuum permitivity, $k_B$ the Boltzmann constant, $T$ the plasma temperature, $n$ the number density, and $q$ the plasma particle charge. A system of charged species is considered a plasma when its Debye length is smaller than its size. The Coulomb coupling parameter is the ratio of the electrostatic energy to the thermal energy of a plasma \cite{fortov} and is defined as

\begin{equation}
    \Gamma = \frac{q^2 / 4\pi \varepsilon_0 r_{WS}}{k_B T}
    \label{eq:gamma}
\end{equation}

Here, $r_{WS}$ is the Wigner-Seitz radius, which can be derived from the number density with $r_{WS} = \sqrt[3]{3/(4 \pi n)}$. A plasma is strongly coupled when its Coulomb coupling parameter $\Gamma>1$.

Cold-plasma experiments can be roughly divided into neutral and non-neutral plasma studies. In neutral plasma, the electron component remains mixed in with the ion component. Recently there have been advances made with neutral plasmas in plasma laser cooling \cite{langin19}, expansion \cite{murphy14, morrison15, forest18}, pair correlations \cite{lyon17, murillo07}, dual-species ion collisions \cite{sprenkle19, boella16}, Rydberg atom-plasma interactions \cite{crockett18}, field-sensing applications \cite{ma17, anderson17}, and quenched randomness and localization \cite{sous19}. Non-neutral plasmas, on the other hand, are composed of only electrons or only ions. Cylindrical non-neutral plasma fluid dynamics have been theoretically analyzed \cite{gould95}, and recent work has been published on dynamical bunching \cite{zerbe18} and Bernstein modes \cite{walsh18}. Work has also been done on expansion of spherically-arranged non-neutral plasma \cite{feldbaum02, kaplan03, bychenkov05, grech11}.

Plasma dynamics are driven by microscopic fields between individual particles and macroscopic fields due to the average space-charge distribution (averaged over a distance on the order of several inter-particle separations). In the present work, we study inter-particle correlations that develop in expanding non-neutral, strongly-coupled plasmas due to micro-field effects, and coarse-scale expansion and shock fronts due to macroscopic fields and external focusing fields. We start with rubidium atoms cooled and trapped in a magneto-optic trap (MOT). From here, plasmas are prepared by photo-exciting a cylindrical region of atoms to an intermediate state, and then ionizing this region with a pulsed ultraviolet laser. We then use direct spatial imaging of the ions to observe the plasma's structure and dynamics. By varying the delay time between plasma formation and imaging, we are able to investigate the plasma's evolution over time. Additionally, we find that varying the plasma density by changing the density of the initial MOT cloud allows us to access qualitatively different domains that lead to different physical effects. In our higher-density plasmas, the dynamics are largely driven by hydrodynamic expansion and external fields, which lead to shock fronts and ion lensing, respectively. Our lower-density plasmas, on the other hand, allow us to observe nearest-neighbor ion correlations, which develop in the course of the time evolution of the strongly-coupled expanding plasma.

Central to our work is the ability to directly image plasma ions with electric fields and spatially-resolving, single-ion-counting micro-channel plate detectors. Ion imaging techniques have been used before on Rydberg atoms in order to observe blockade radii \cite{schwarz13}, van der Waals interactions \cite{thai15}, ionization spectra \cite{grimmel19, stecker19}  and tunnel ionization rates \cite{gawlas19}.

\section{Experimental Setup} \label{sec:setup}

We employ a two-MOT vacuum chamber with a primary MOT that contains a reservoir of trapped rubidium-87 atoms, and a secondary MOT from which the plasma is formed. Atoms are loaded from the primary MOT into the secondary with a pulsed pusher laser (1.5 mW peak power, 1 mm FWHM beam diameter, and 10 Hz repetition rate with duty cycle of 10\%). When the pusher laser is at maximum power, the secondary MOT is able to trap clouds of up to $\sim 10^7$ \textsuperscript{87}Rb atoms at densities of up to $\sim 4 \times 10^{10}$~cm$^{-3}$. We can decrease the density and extent of the secondary-MOT atom cloud in a controlled manner by decreasing the power of the pusher laser. The ionization region is electric-field-zeroed to within 10~mV/cm using internal electrodes to allow us to study the unperturbed expansion of the plasma.

The experiment is run at a repetition rate of 10 Hz. In each cycle, a plasma is formed in the secondary MOT with a two-photon photoionization process. Prior to plasma formation, the MOT light is turned off to prevent unwanted plasma outside the intended initial volume. A fraction of the cold atoms is driven by an 18 $\mu$s long, 780-nm laser pulse to the 5$P_{3/2}$ state. This laser beam has a Gaussian profile, with a waist $w_0$ of 9 microns, Rayleigh range of 326 microns, and central intensity of 105\textit{I\textsubscript{sat}}. Five microseconds after the 780-nm pulse is turned on, the 5$P_{3/2}$ atoms are ionized with a 10 ns, 335-nm ultraviolet pulse from a Q-switched, frequency-tripled Nd:YAG laser. This wavelength is well above the ionization threshold wavelength of 479.1 nm for atoms in the 5$P_{3/2}$ state. Atoms in the intermediate state become ionized, and the liberated valence electrons have a kinetic energy of 0.9 eV, which is sufficient to escape the plasma cloud. The initial temperature of the trapped atoms is roughly 100~$\mu$K, but the photoionization is accompanied with recoil heating to $\approx$44~mK, the initial temperature of our plasmas. The 355-nm pulse has a diameter of $\sim 2$~mm, which is much larger than the size of the 780-nm beam. Hence, the initial geometry of the plasma column is determined by the 780-nm beam size and its central intensity, and the MOT size. Further, the fluence of the 355-nm pulse is $\lesssim 10^{16}$~cm\textsuperscript{-2}, corresponding to a photoionization probability of the 5$P_{3/2}$ atoms of $\lesssim 10$ percent, leading to up to several hundred ions in the initial plasma volume.

\begin{figure}
    \centering
    \includegraphics[width = 8 cm]{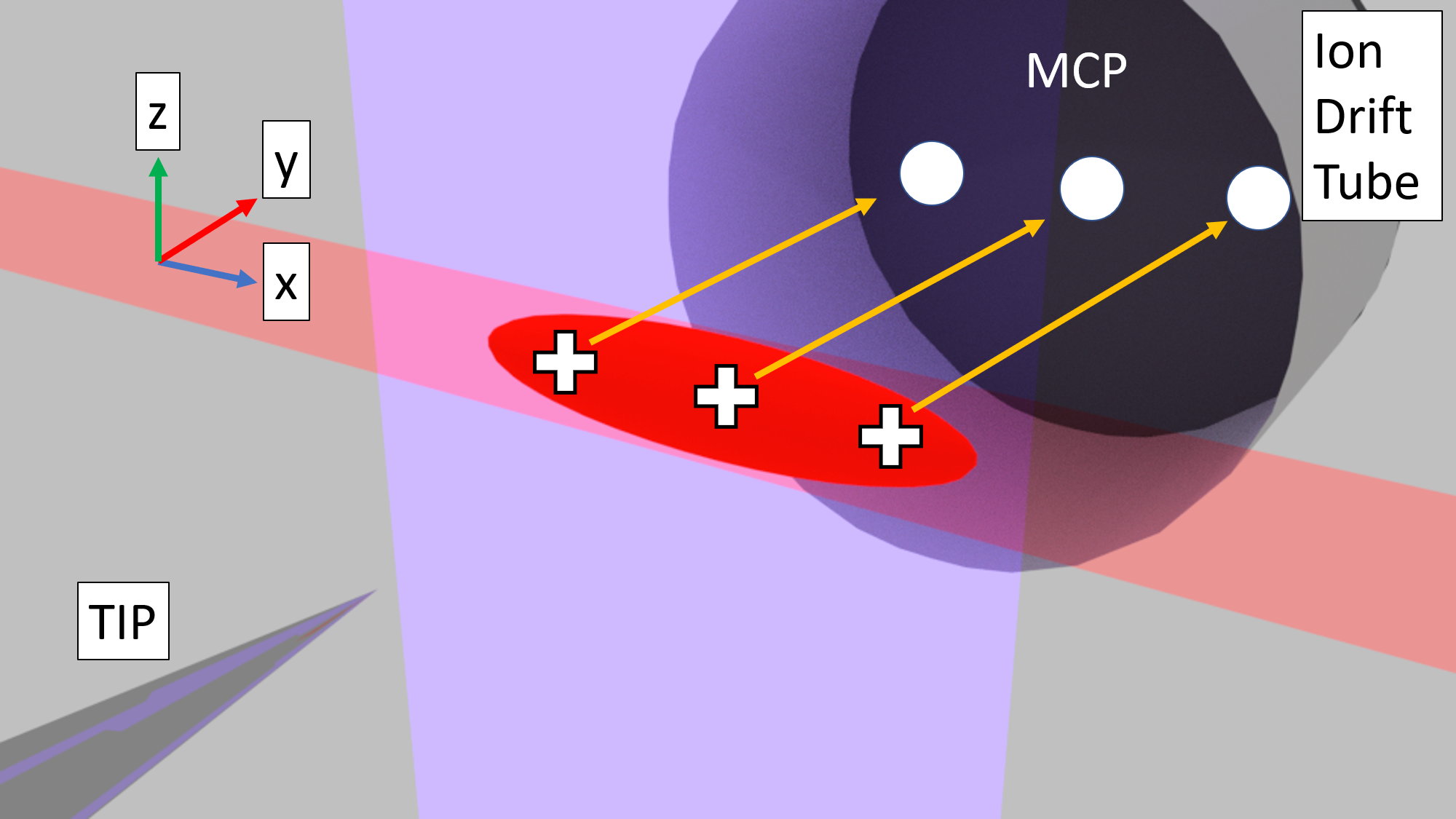}
    \caption{Artistic representation of the plasma formation and imaging process. The positively charged rubidium ions are pushed by the TIP electric field through the ion drift tube to the MCP, where they produce white blips that can be imaged. This image is not to scale.}
    \label{fig:config}
\end{figure}

The MOT vacuum chamber contains a needle-shaped, beryllium-copper tip imaging probe (TIP), which is positioned $\sim 2$ mm away from the excitation region. Figure~\ref{fig:config} shows the orientation of the plasma, TIP, and other experimental components. After the plasma has been formed, it is allowed to expand for a variable wait time $\tau$. After the wait time, a high voltage step pulse with a 90\% rise time of 76 ns is applied to the TIP, producing a divergent strong electric field. The ions are accelerated by this electric field towards a microchannel plate (MCP). Upon impact on the MCP, the ions produce bright spots on a phosphor screen which give the plasma ion positions at the time of extraction. The MCP detection efficiency is 30-50\%, according to manufacturer's specifications. A CCD camera takes a picture of the phosphor screen for each experimental cycle. The single-ion resolving images represent the plasma ion distributions projected onto a plane transverse to the extraction trajectory, taken at time $\tau$. The spatial structure of the expanded plasma along the extraction trajectory is analyzed by measuring its time-of-flight distribution to the MCP using a multichannel scaler (SRS Model SR430). The magnification factor of the imaging setup can be varied by adjusting the distance between the TIP and the excitation region and can be calibrated by translating the excitation region by known distances.

\section{Modeling} \label{sec:model}

We use trajectory simulations and a fluid description to model the experimental data. In the trajectory simulations we use up to 1000 ions and account for all external and inter-particle Coulomb forces. A numerically generated map of the electrostatic potential along the ion-imaging path between the TIP and the MCP is implemented to model the ion imaging. The timing in the simulation program, as well as the analysis of the ion images on the MCP, the ion time-of-flight data, and the pair correlation functions closely follows the experimental procedure. The trajectory simulation is exact, because it accounts for microscopic and macroscopic fields, initial ion temperature, and Coulomb collisions during free expansion, ion extraction and imaging. Hence, the trajectory simulations are well-suited to model correlations and strong-coupling effects, in addition to modeling the overall expansion and imaging dynamics.

In the fluid model, the plasma is described as an infinitely long, cylindrical, continuous space charge distribution with a radial profile given by the ion excitation parameters used in the experiment. The space charge is then treated as a discrete set of thin cylindrical shells with given charge and mass. The (discrete) radii of the shells are then propagated using Newton's equations.  Gauss's law is employed to track the macroscopic electric field at the locations of the shells. The dynamical variables in this model are the shell radii and velocities, and the electric field. These data also allow the generation of plasma density maps versus time, with the microscopic fields turned off. The fluid model is well-suited to model shock fronts, which appear as singularities of the fluid-model plasma density. In the fluid model, the effects of the initial photon-recoil-induced ion motion are ignored.

\section{Higher-Density Plasma Regime} \label{sec:highdensity}

The higher-density plasma is excited from an atom cloud containing $3.8 \times 10^7$ atoms at a density of $4.1 \times 10^{11}$ atoms/cm$^3$. The initial ion density is $n=2.4 \times 10^8$ cm$^{-3}$. We observe this plasma at $\tau = 0$, 2, 4, 6, 8, and 10 microseconds, and we run for 5000 cycles at each of these expansion times. The magnification factor of the imaging for this data set is 54 times. The initial Debye length is $\lambda_D=0.94$ $\mu$m, and the initial Coulomb coupling parameter is $\Gamma=39$.

\begin{figure}
    \centering
    \includegraphics[scale=0.4]{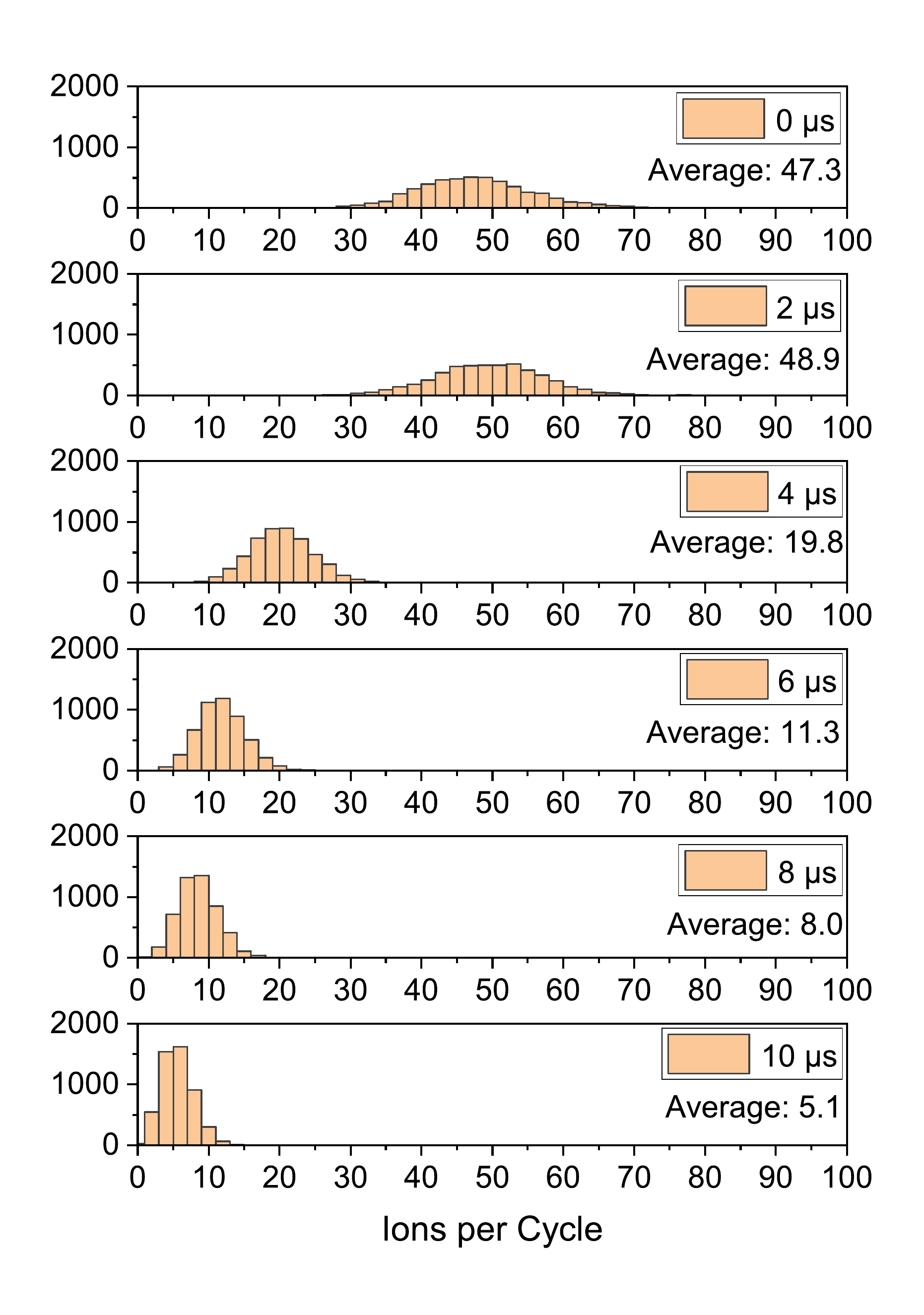}
    \caption{Histograms of ion counts per cycle in the higher-density regime.}
    \label{fig:hist_high}
\end{figure}

\begin{figure}
    \centering
    \includegraphics[width=4 cm]{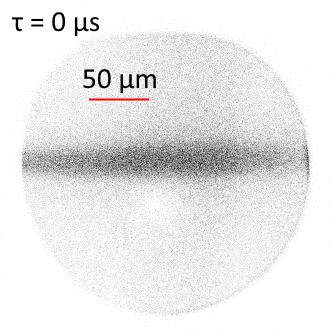}
    \includegraphics[width=4 cm]{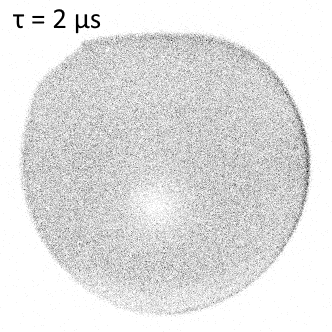}
    \includegraphics[width=4 cm]{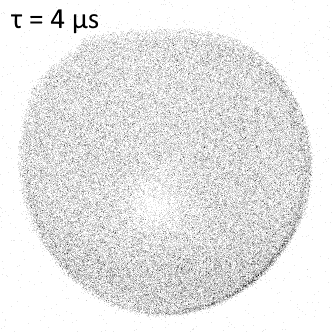}
    \includegraphics[width=4 cm]{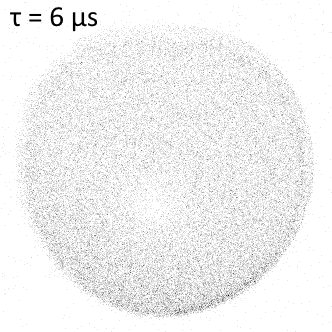}
    \includegraphics[width=4 cm]{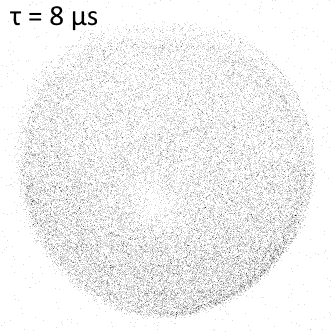}
    \includegraphics[width=4 cm]{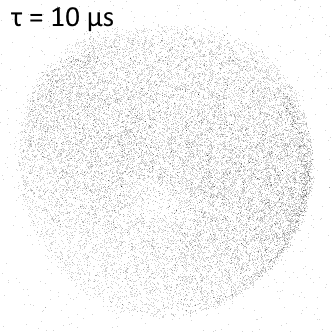}
    \caption{Averaged pictures of higher-density plasmas for the indicated expansion times. The distance scale in the upper-left image corresponds to distance in the object plane. The light spot is an area of reduced ion detection rate; its origin is under investigation.}
    \label{fig:pics_high}
\end{figure}

\begin{figure}
    \centering
    \subfloat[]{
        \includegraphics[width=4 cm]{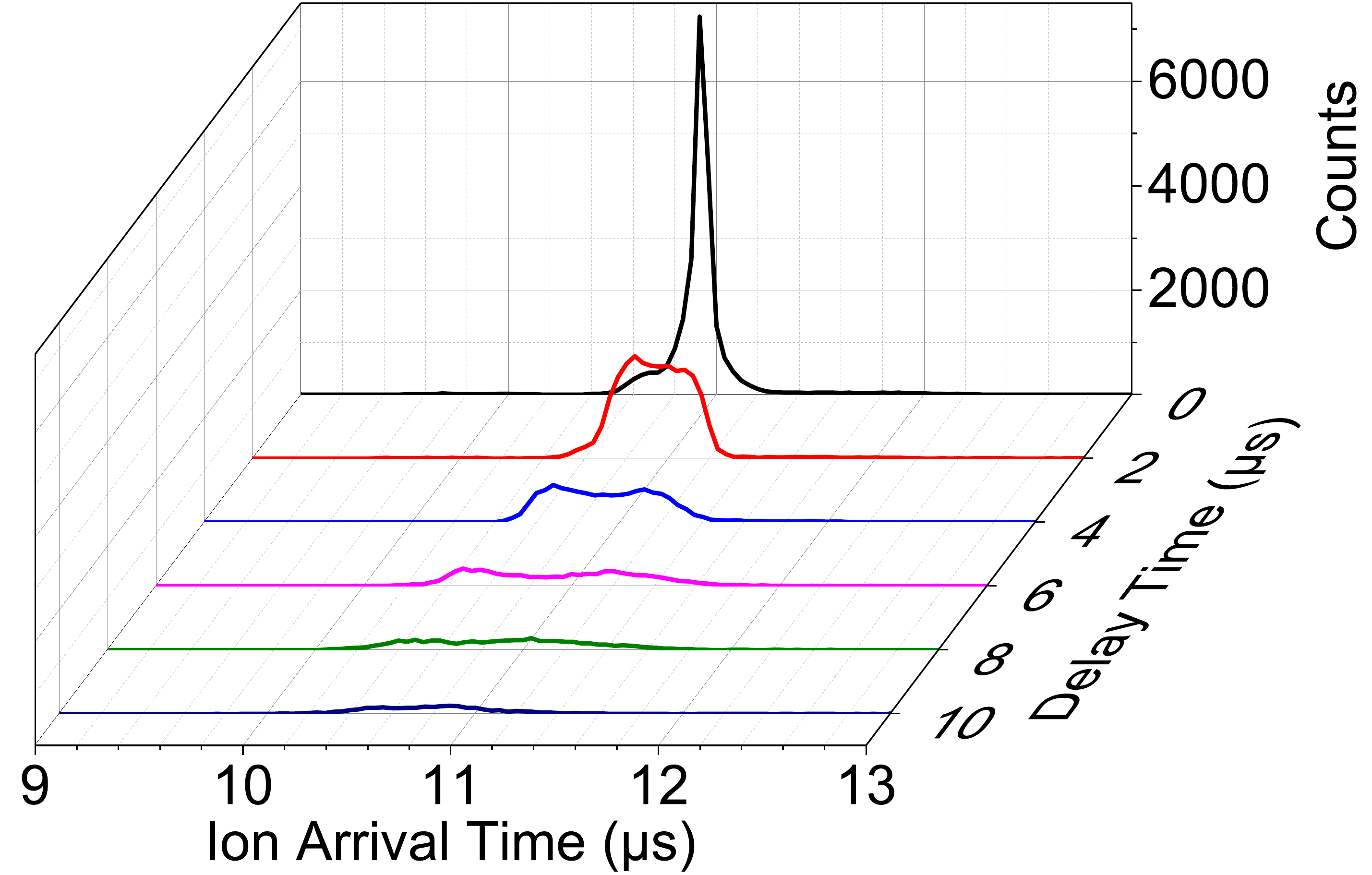}
        \label{subfig:multi_high_exp_lin}
        }
    \subfloat[]{
        \includegraphics[width=4 cm]{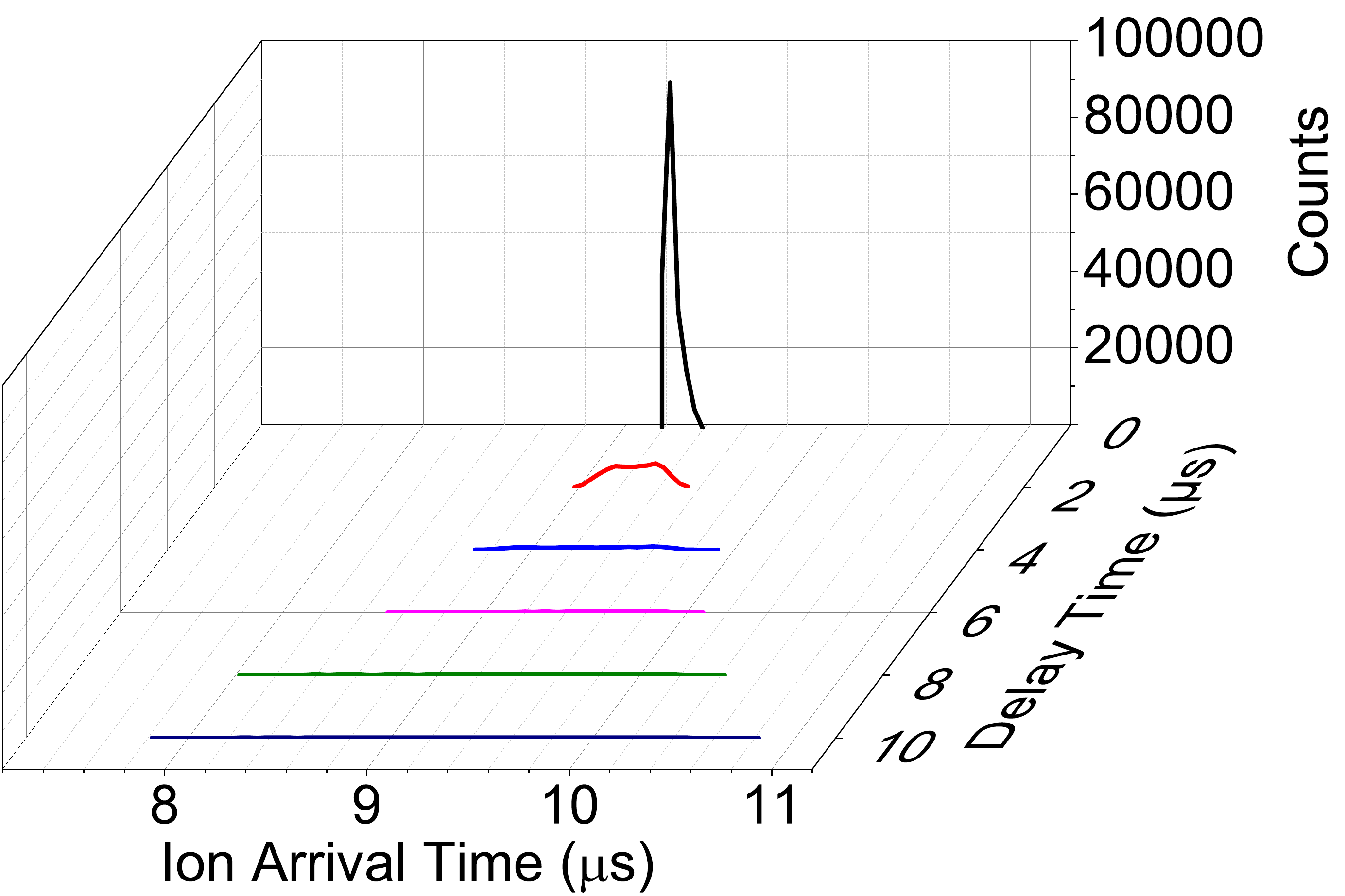}
        \label{subfig:multi_high_comp_lin}
        }
    \linebreak
    \subfloat[]{
        \includegraphics[width=4 cm]{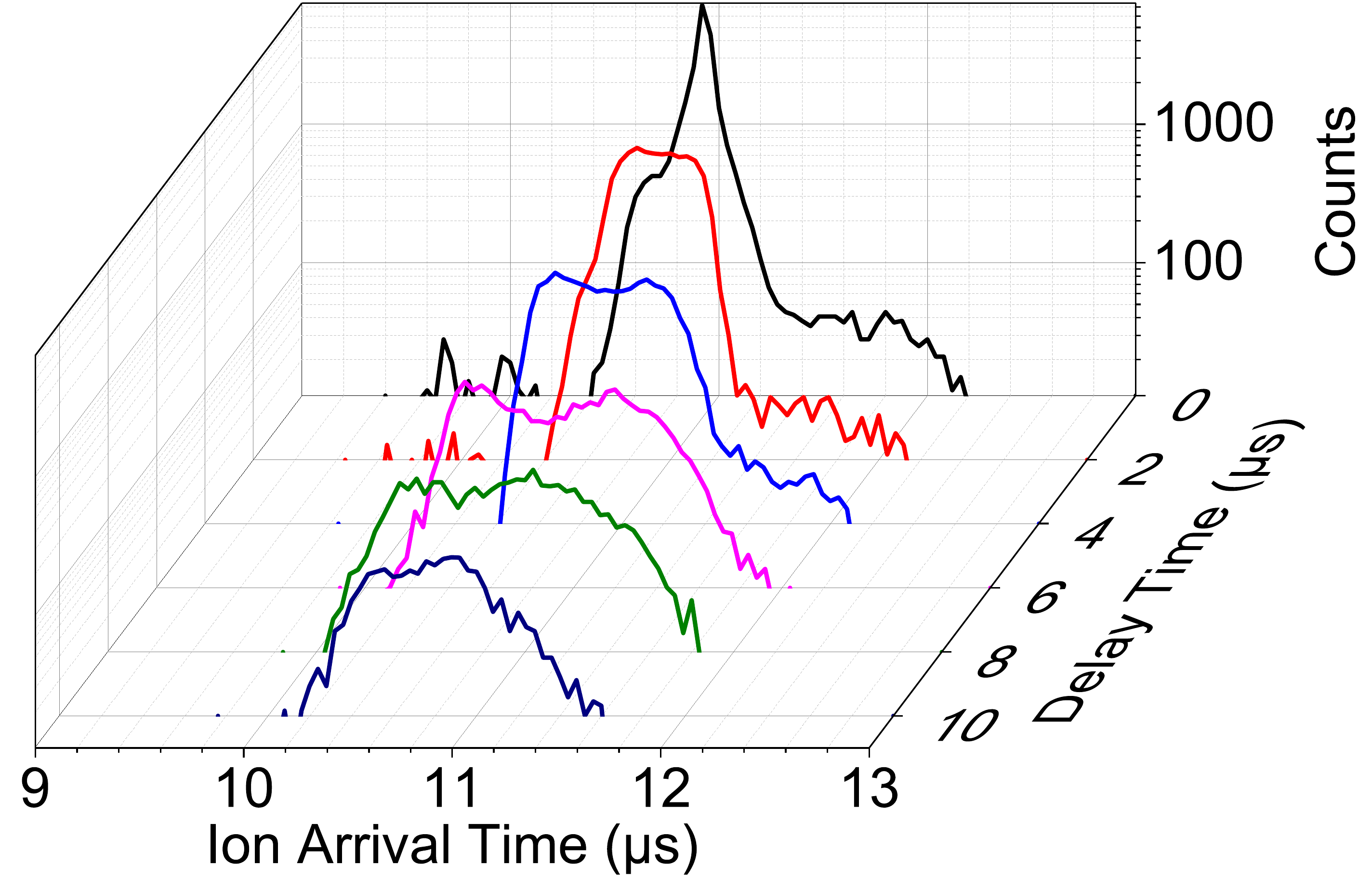}
        \label{subfig:multi_high_exp_log}
        }
    \subfloat[]{
        \includegraphics[width=4 cm]{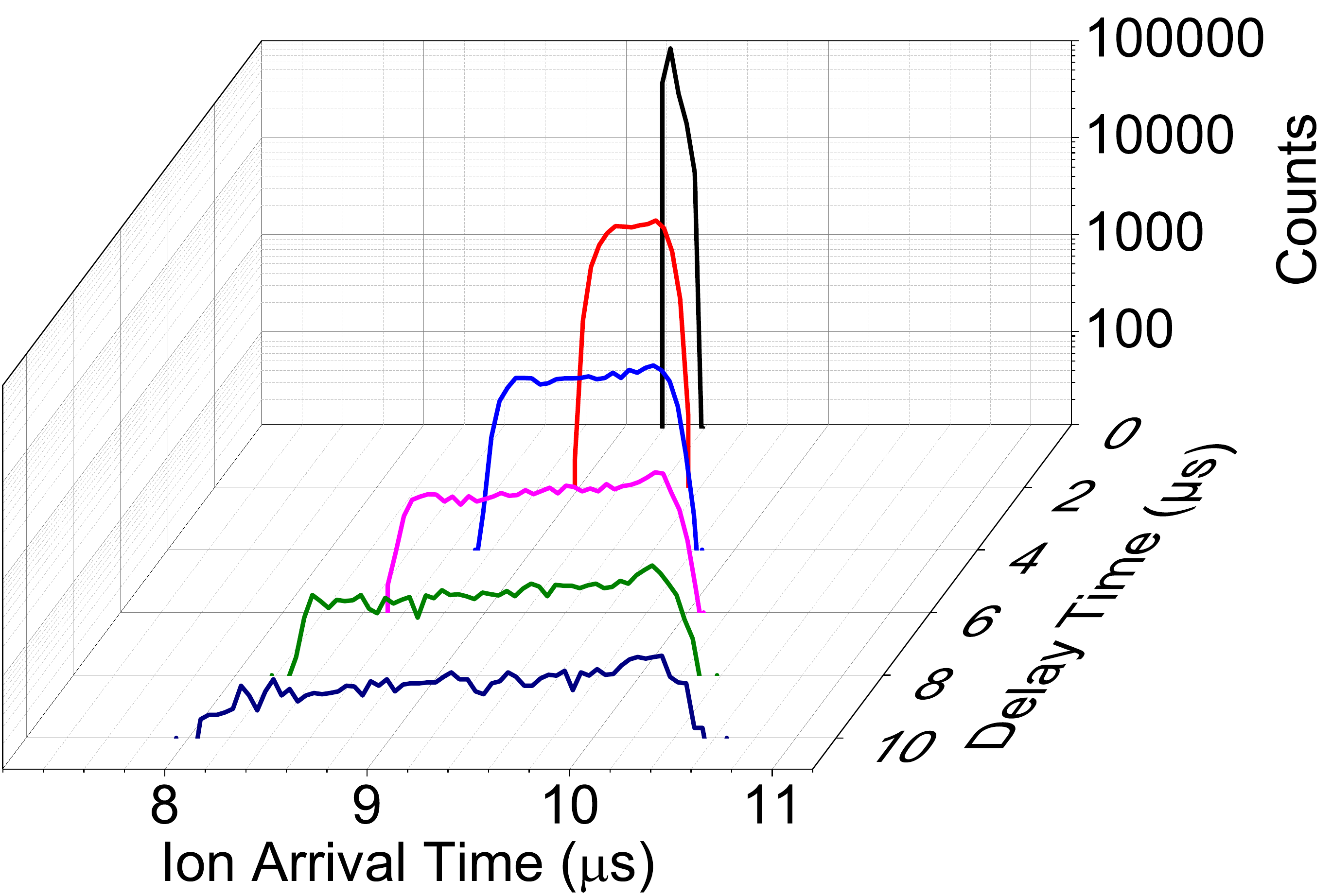}
        \label{subfig:multi_high_comp_log}
        }
    \caption{Experimental (\ref{subfig:multi_high_exp_lin} and \ref{subfig:multi_high_exp_log}) and computational (\ref{subfig:multi_high_comp_lin} and \ref{subfig:multi_high_comp_log}) distributions of ion arrival times (horizontal axis) for different plasma expansion times (tilted axis) at higher density, shown on both linear (top) and logarithmic (bottom) scales. The linear scale is useful for visualizing overall ion counts between traces, while the logarithmic scale is better for displaying faint signals at longer expansion times.}
    \label{fig:multi_high}
\end{figure}

Figures~\ref{fig:hist_high}~and~\ref{fig:pics_high} show the ion count statistics and averaged images of plasma expansion. At $\tau = 2$ $\mu$s, the plasma has expanded to fill the field of view, with the average ion count remaining approximately constant. For longer expansion times, the ions leave the field of view, and the average ion count monotonically decreases. Figure~\ref{fig:multi_high} shows the arrival time distributions of the plasma ions for each expansion time. The arrival time traces have been aligned on the time axis so that the firing of the TIP voltage pulse is at $t=0$. As the plasma expansion time increases, the ion arrival time distribution bifurcates into a leading and a lagging peak, signaling the formation of shock fronts. As the plasma expands radially, a fraction of ions move toward the TIP, while another fraction of ions move in the opposite direction toward the MCP. When the TIP electric field is engaged, the ions closer to the TIP experience a much stronger electric field than ions that are farther away. Hence, the longitudinal position distribution of the ions maps onto a time-of-flight distribution, with shorter times of flight corresponding to ions closer to the TIP. The peaks in the time-of-flight distribution correspond to shells of enhanced plasma density, or shock fronts. In our data in Fig.~\ref{fig:multi_high}, the shock fronts begin to form at $\tau = 4$ $\mu$s and are most prominent at $\tau = 6$ $\mu$s.

\begin{figure}
    \centering
    \includegraphics[width = 8 cm]{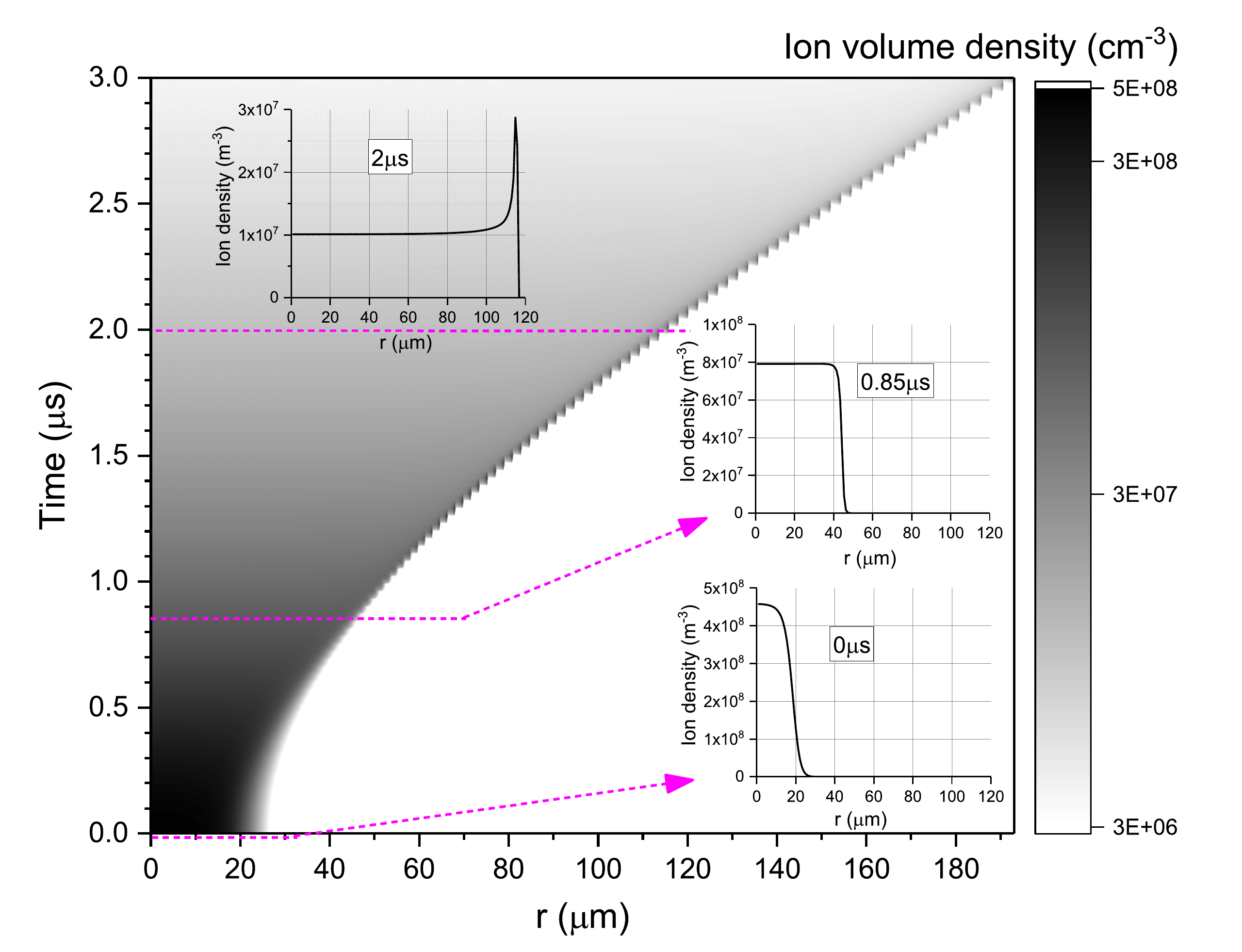}
    \caption{Fluid-model simulation of the expansion of an azimuthally symmetric plasma. Plasma density is shown vs radius (horizontal) and expansion time (vertical axis). A shock front develops at $0.85~\mu$s and becomes increasingly pronounced thereafter.}
    \label{fig:fluid}
\end{figure}

The fluid model described in Sec.~\ref{sec:model} is well suited to treat the formation of shock fronts. In Fig.~\ref{fig:fluid} we show a fluid-model simulation of the expansion of an azimuthally symmetric, infinitely long plasma with an initial diameter of $\approx 50~\mu$m, an initial saturated Gaussian profile, and a total linear charge density of $5 \times 10^{5}$~e/m. This corresponds with the highest-density plasmas we have studied in our experiment. The figure shows that the shock front forms at about 0.85~$\mu$s expansion time (see insets in Fig.~\ref{fig:fluid}). In the fluid model, the onset of the shock front occurs when initially further-inside charged shells overtake initially further-outside charged shells. The radial location of the shock front is given by the condition $d r_i (t)/ d i = 0$, where $r_i(t)$ is the radius of the $i$-th shell at time $t$, and $i$ is an (integer) counter that counts the charged shells from the inside out (at time $t=0$). According to this equation, the singular behavior, which is equivalent to the shock front, marks a condition where a group of shells, with indices $i$ within a contiguous range, pile up at approximately the same radius, which is the shock-front location. This can only occur after some expansion time. At later times, the shock front becomes more pronounced, while in the interior region the plasma density converges to a constant.

Comparing trajectory and fluid models, it is noted that the shock fronts are washed out due to the effects of the micro-fields and the initial recoil-induced velocity distribution of the ions. This can be seen, for instance,
in snapshots of the
trajectory-model results, where instead of singular ion-density enhancements near the ion-cloud edges one finds ion-density enhancements on the order of two (for our conditions). Another evidence of the micro-field effects is that the shock-front signatures in Fig.~3, where we show both experimental and trajectory-model results, are moderate signal enhancements, as opposed to singular spikes, at the beginning and the end of the time-of-flight distributions.

\begin{figure}
    \centering
    \subfloat[]{
        \includegraphics[width=4 cm]{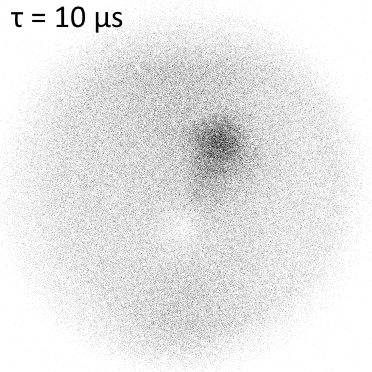}
        \label{subfig:quadrupole_exp}
        }
    \subfloat[]{
        \includegraphics[width=4 cm]{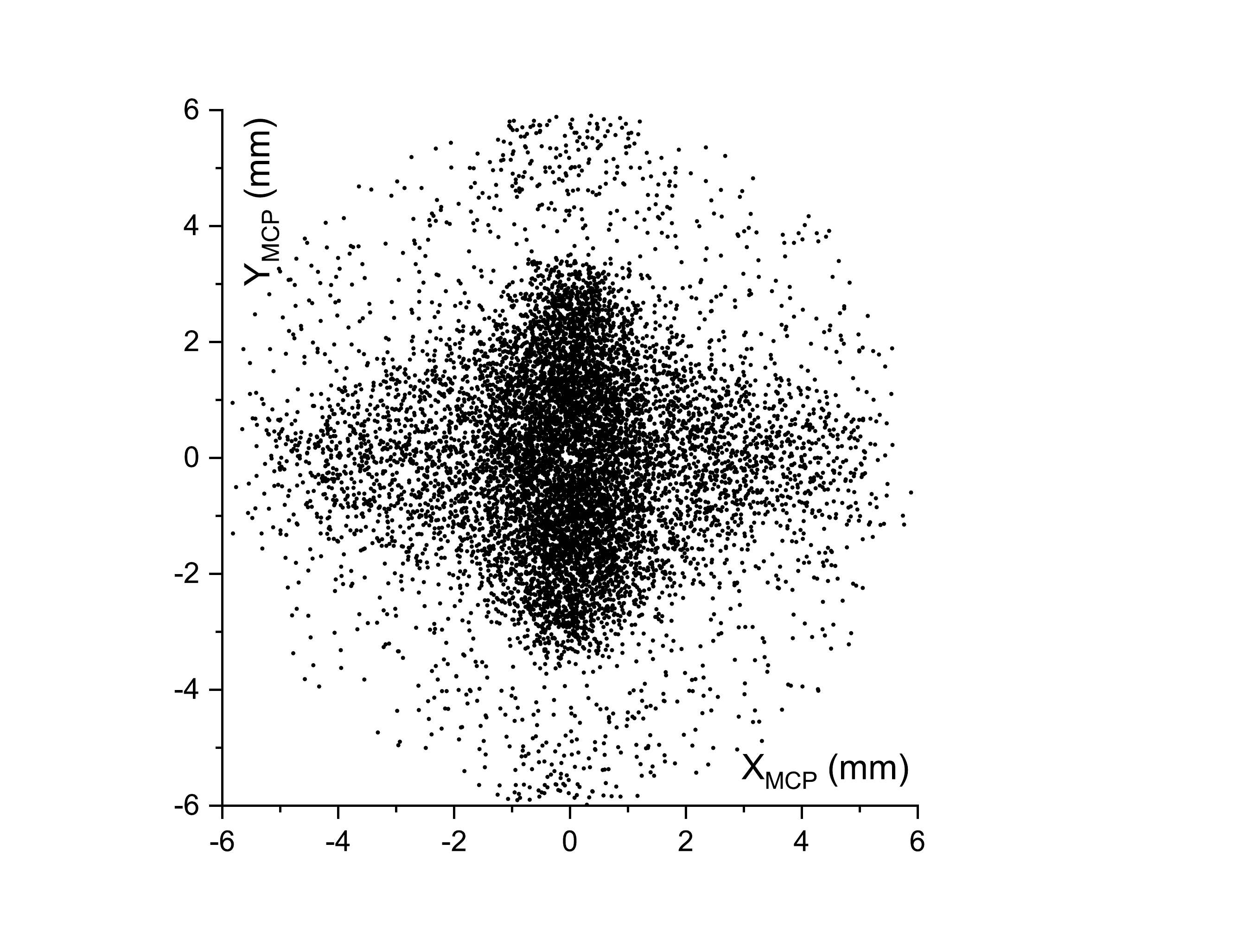}
        \label{subfig:quadrupole_comp}
        }
    \caption{Experimental (\ref{subfig:quadrupole_exp}) and computational (\ref{subfig:quadrupole_comp}) MCP images of higher-density plasma at $\tau = 10$ $\mu$s in the presence of a quadrupole electric field. Instead of expanding outwards and leaving the field of view, the ions are refocused by the electric field and accumulate on a localized region of the MCP.}
    \label{fig:quadupole}
\end{figure}

In the course of our data collection, we have noted that, to study free expansion, it is not sufficient to zero the electric field at the location of the initial plasma cloud. It is also necessary to also field-zero the linear (quadrupole) components of the electric field. Incomplete field-zeroing results in a quadrupole field centered at the ionization region. This field can cause the expanding plasma to refocus along certain directions of space, while defocusing along other (orthogonal) directions. Figure~\ref{fig:quadupole} shows experimental and computational MCP images of a higher-density plasma at $\tau = 10$ $\mu$s in the presence of a quadrupole field that focuses the ions in the $xz$-plane and defocuses along the $y$-direction (coordinates defined in Fig.~\ref{fig:config}). In this case, the plasma focus generated by the quadrupole field manifests in a localized, kite-shaped region on the MCP of enhanced ion count density. The kite shape arises from astigmatism produced by the lack of azimuthal symmetry of the initial plasma cloud along the $y$-direction (see Fig.~\ref{fig:config}). In the present work, we have thoroughly field-zeroed to avoid this plasma focus effect.

\section{Lower-Density Plasma Regime} \label{sec:lowdensity}

The lower-density plasma is prepared from an atom cloud containing $1.3 \times 10^6$ atoms at a density of $7.7 \times 10^{10}$ atoms/cm$^3$. The initial ion density is $n=9.3 \times 10^7$ cm$^{-3}$, Due to the smaller number of counts, compared to the higher-density plasma case in Sec.~\ref{sec:highdensity}, in the lower-density case we limit the expansion time $\tau$ to 3.5 $\mu$s, and we use shorter time steps of 0.5 $\mu$s. The most interesting expansion and correlation dynamics occur over that time scale. Again, we take 5000 shots at each of these expansion times. The magnification factor of the imaging for this data set is 60 times. The initial Debye length is $\lambda_D=1.5$ $\mu$m, and the initial Coulomb coupling parameter is $\Gamma=28$.

\begin{figure}
    \centering
    \includegraphics[scale=0.4]{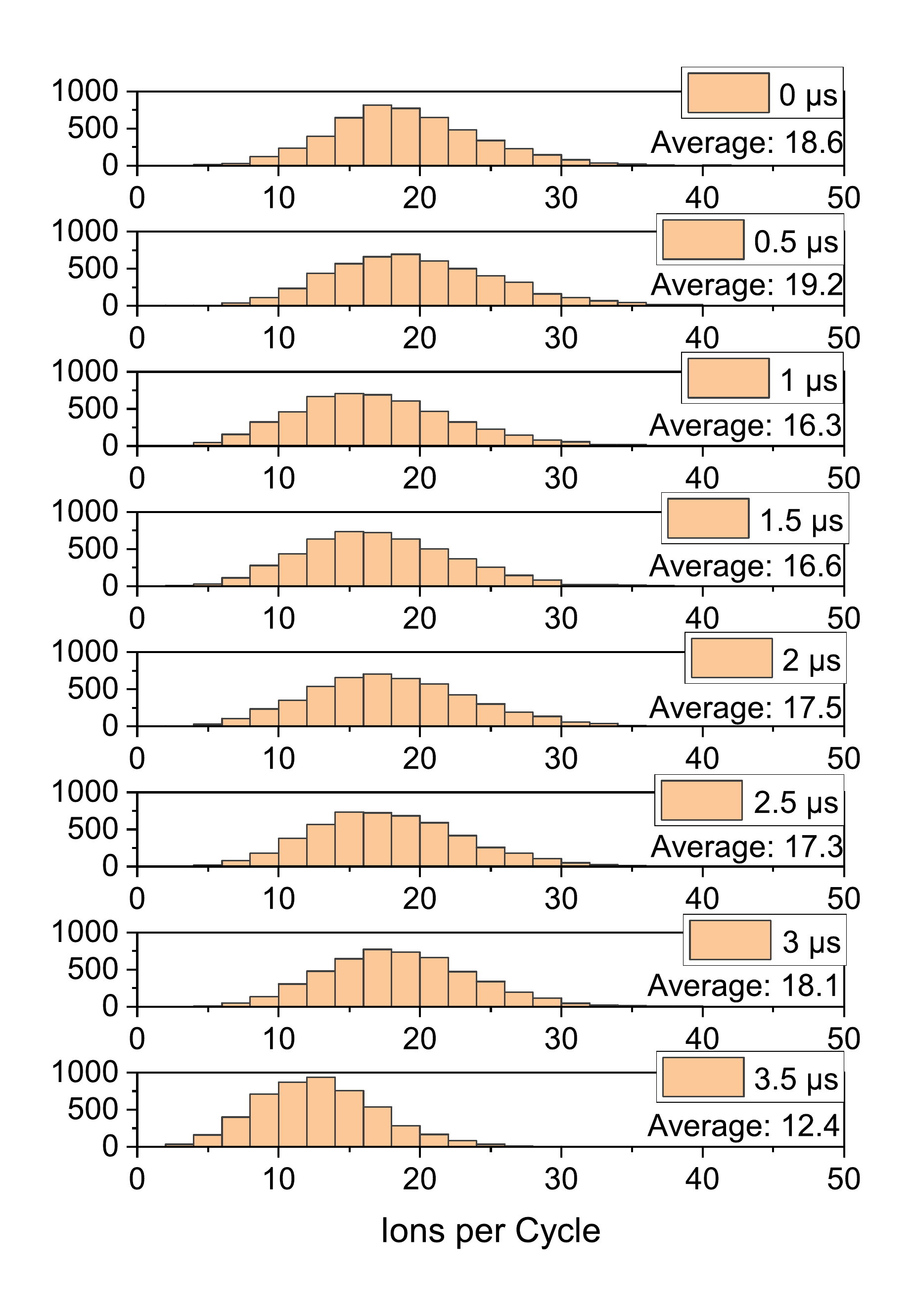}
    \caption{Histograms of ion counts per cycle in the lower-density regime.}
    \label{fig:hist_low}
\end{figure}

\begin{figure}
    \centering
    \subfloat[]{
        \includegraphics[width=4 cm]{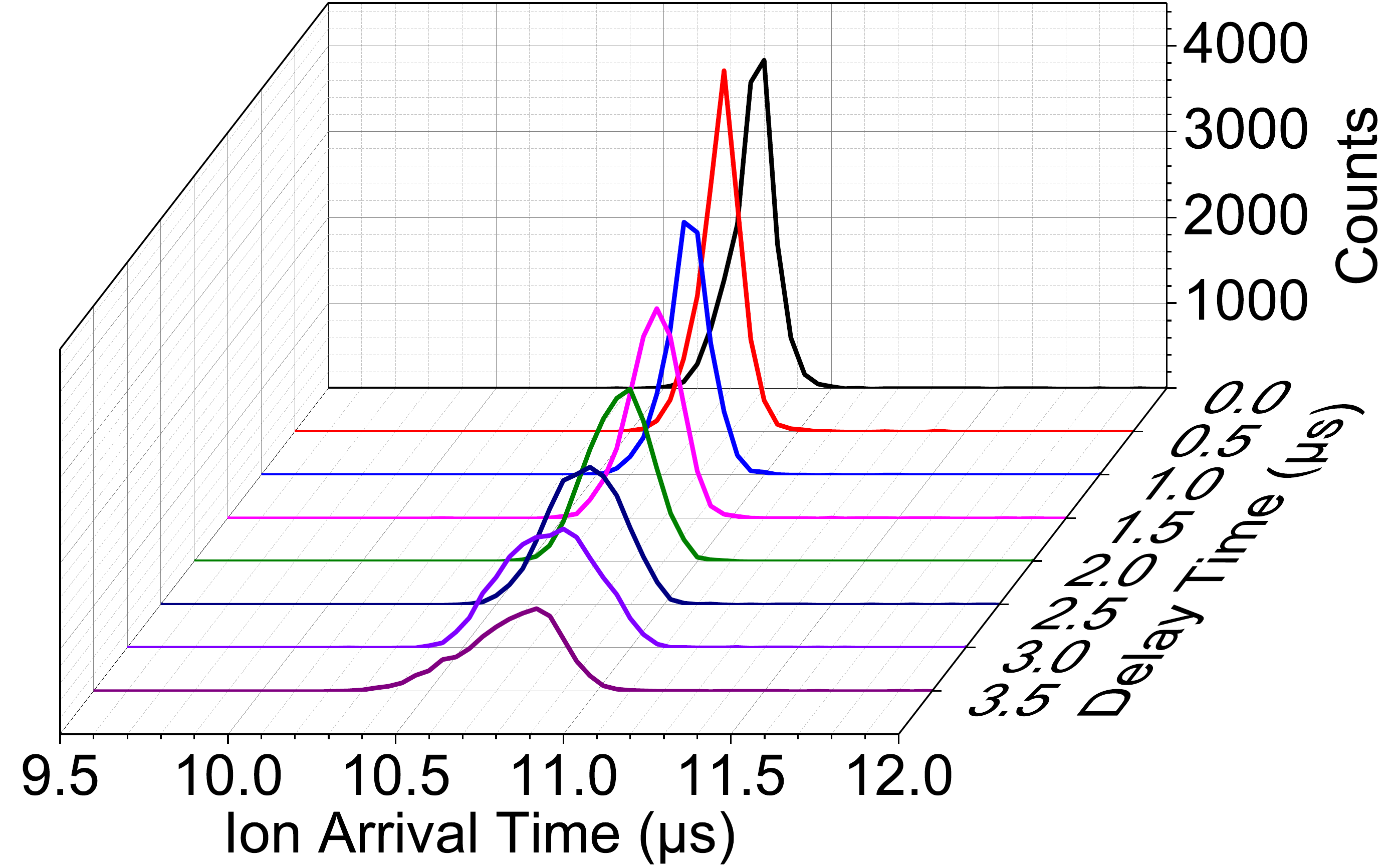}
        \label{subfig:multi_low_exp}
        }
    \subfloat[]{
        \includegraphics[width=4 cm]{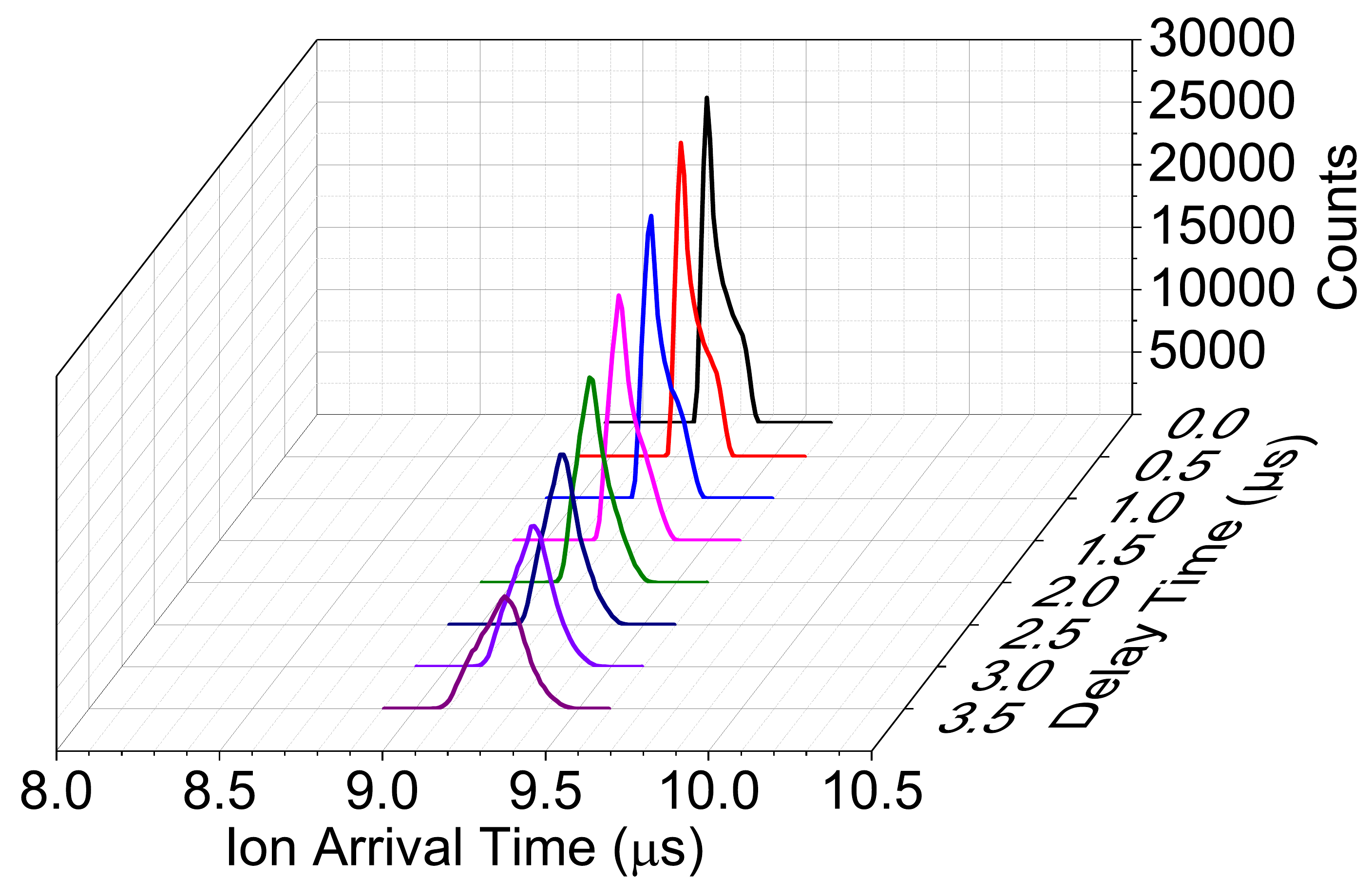}
        \label{subfig:multi_low_comp}
        }
    \caption{Distribution of ion arrival times for different plasma expansion times at lower density.}
    \label{fig:multi_low}
\end{figure}

\begin{figure}
    \centering
    \includegraphics[width=4 cm]{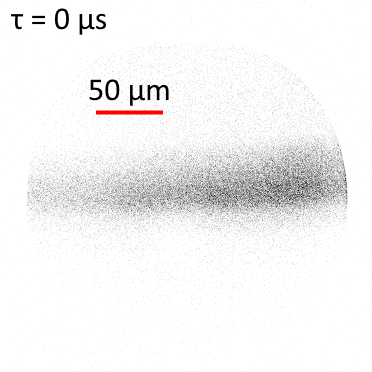}
    \includegraphics[width=4 cm]{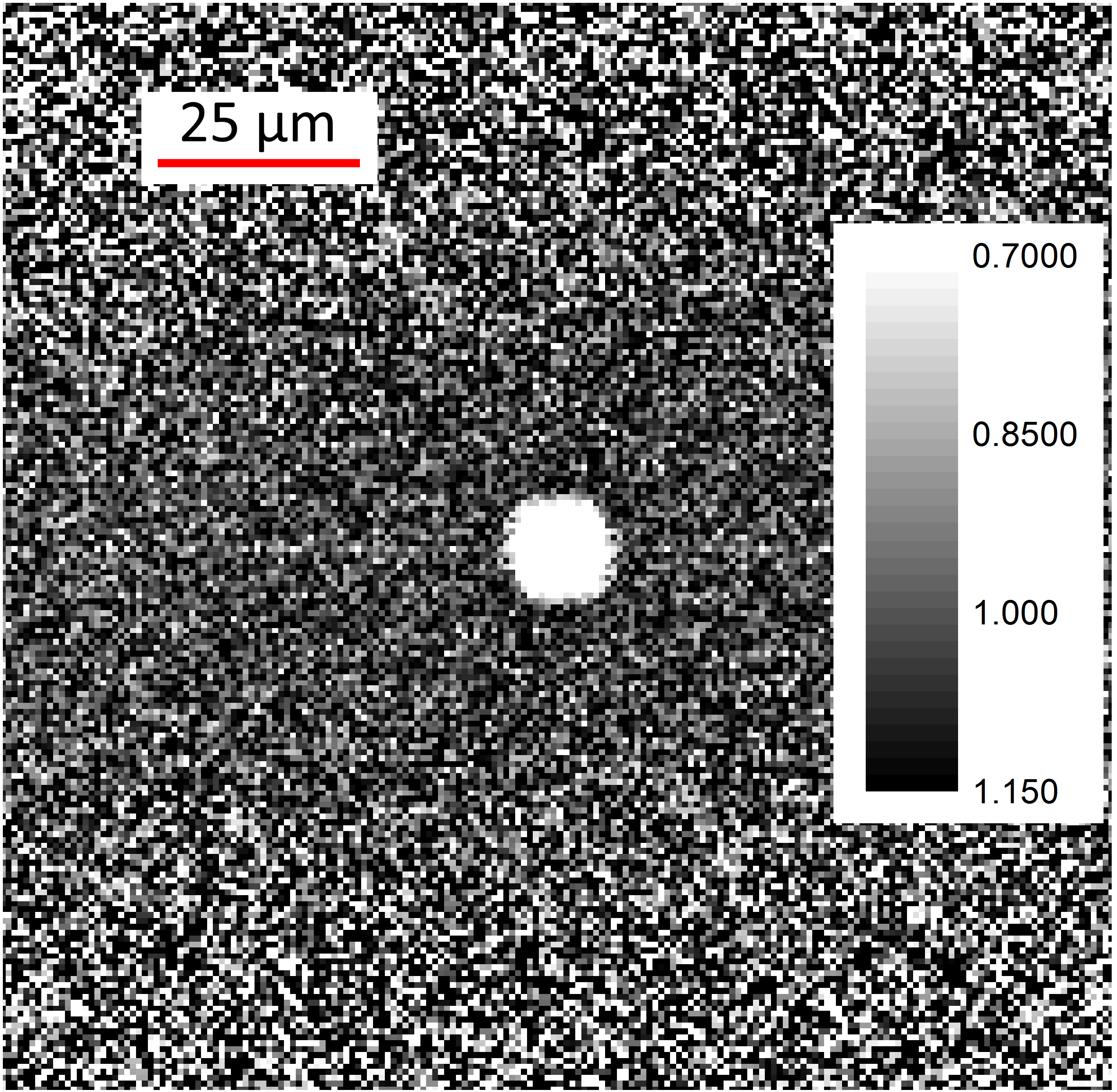}
    \includegraphics[width=4 cm]{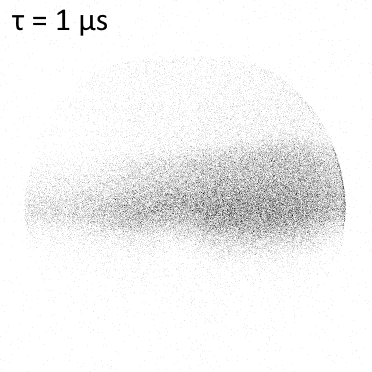}
    \includegraphics[width=4 cm]{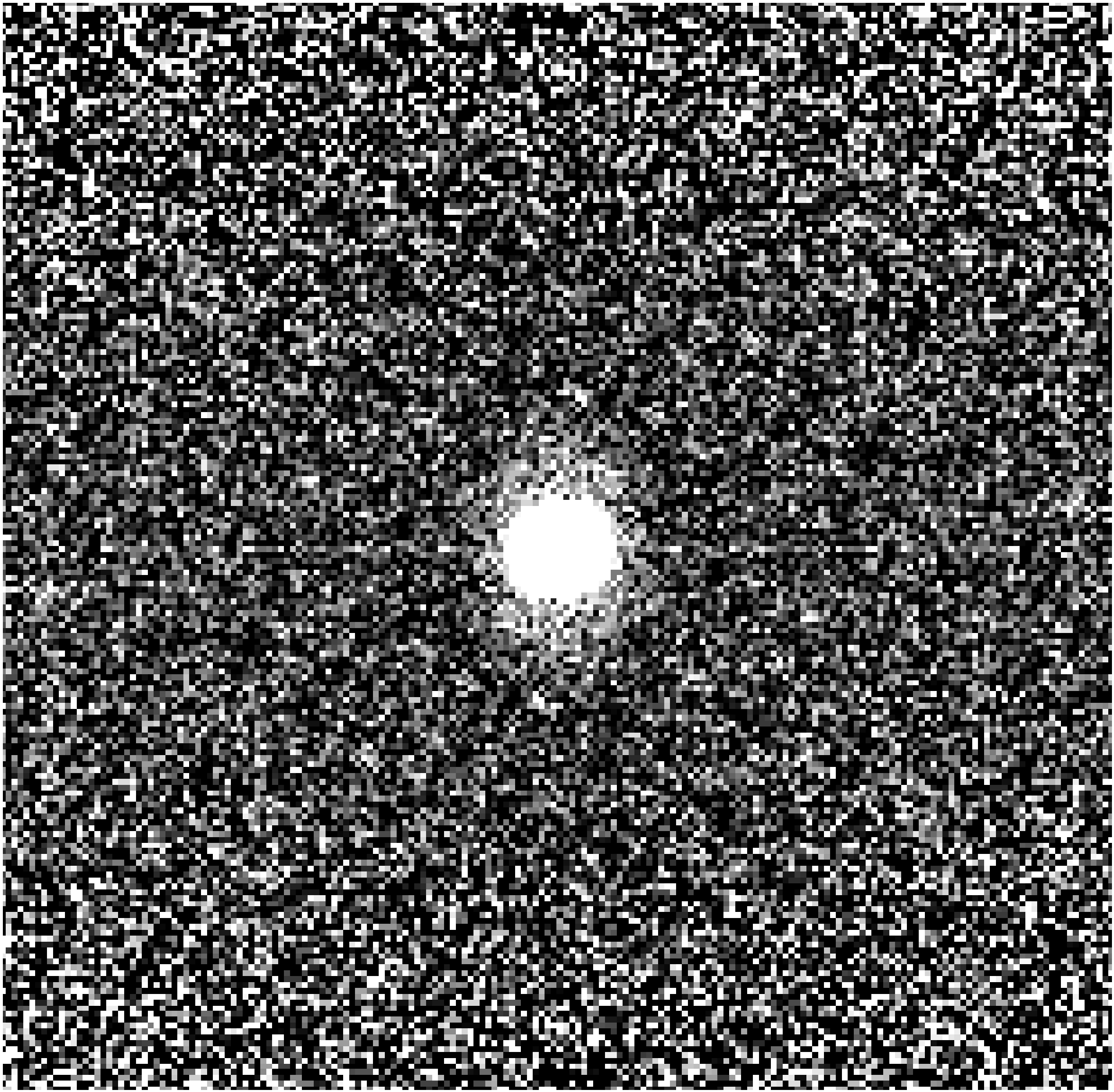}
    \includegraphics[width=4 cm]{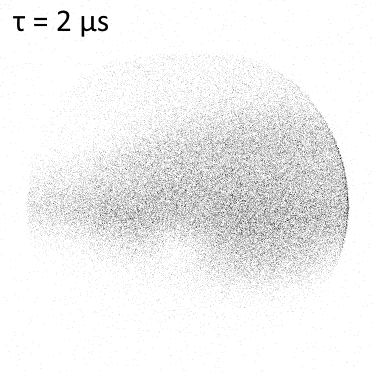}
    \includegraphics[width=4 cm]{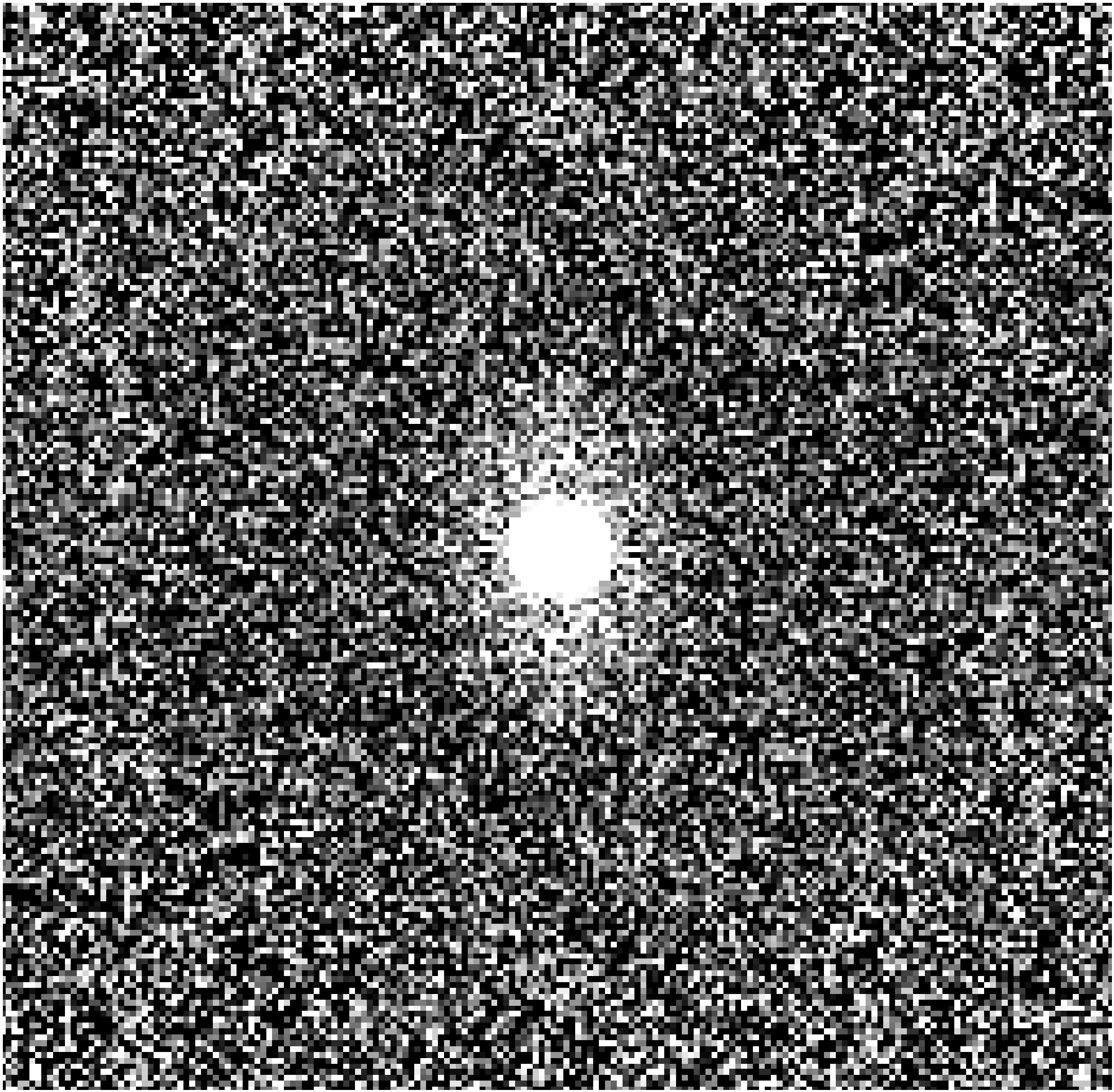}
    \includegraphics[width=4 cm]{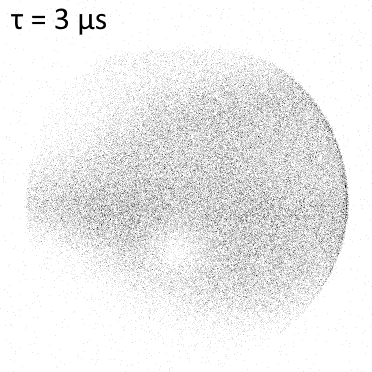}
    \includegraphics[width=4 cm]{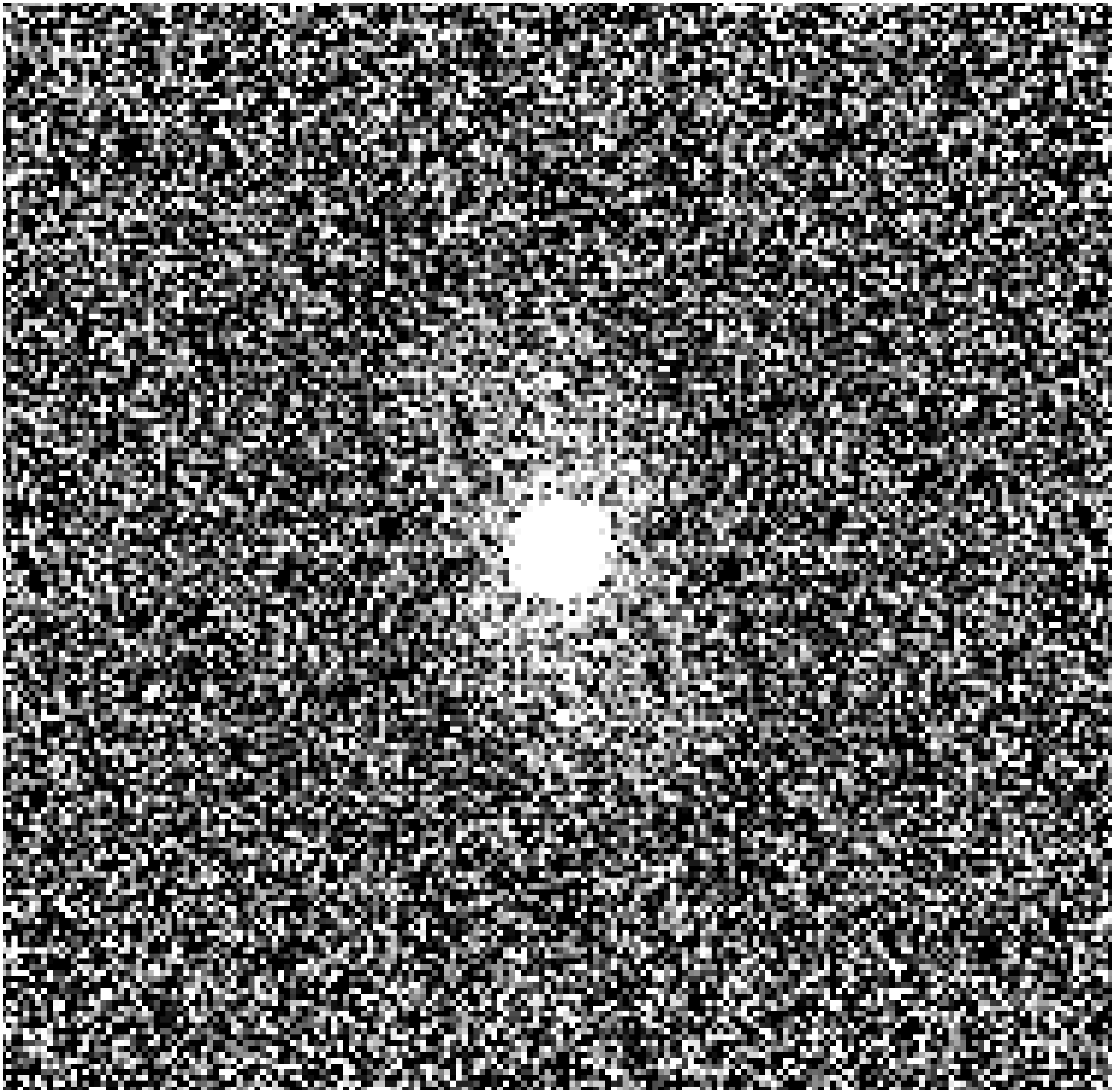}
    \caption{Images of lower-density plasma expansion. On the left are averaged images of expansion at the indicated expansion times $\tau$. On the right are pair correlations of the raw images at the same time steps. The grayscale is calibrated such that a value of 1 implies no correlation.}
    \label{fig:pics_low}
\end{figure}

Figures~\ref{fig:hist_low} and~\ref{fig:multi_low} show the ion count statistics and arrival time distributions. Because this plasma is lower in density and smaller in size, it exhibits a less forceful Coulomb expansion. As a result, the average ion count remains fairly constant for $\tau < 3.5$ $\mu$s; even at $\tau = 3.5$ $\mu$s only a small fraction of ions leaves the field of view (see Figs.~\ref{fig:hist_low} and~\ref{fig:pics_low}). We also no longer observe a bifurcation of ion arrival times, though the arrival time distribution does broaden with longer expansion times. These observations indicate that the lower-density plasma does not produce pronounced shock fronts over the course of its expansion.

The lower-density plasmas are well-suited to study correlation effects that develop on a timescale of several microseconds. During that time, the higher-density plasma studied in Section~\ref{sec:highdensity} Coulomb-expands to a size exceeding the field of view of the MCP. Due to its large size, any correlations that are present within the three-dimensional plasma system become obscured in the projection onto the MCP plane. This is not the case in the lower-density plasma because of its smaller size along the projection direction ($y$-direction in Fig.~\ref{fig:config}), and its reduced density of counts in the MCP plane ($xz$ plane).

We calculate ion pair correlations at each expansion time $\tau$ by first processing the raw images with a peak finder algorithm, obtaining the pair correlation of the peaks, and normalizing the  repetition-averaged pair correlation such that a value of one corresponds to an absence of correlations. In Fig.~\ref{fig:pics_low}, we show images that are averages over the 5000 repetitions for each delay time, along with the corresponding normalized pair correlations. On a macroscale (distances $\gtrsim$ 100 $\mu$m), the pair-correlation functions expand in a manner that reflects the overall plasma expansion. This is seen in Fig.~\ref{fig:pics_low} for $\tau \leq 1$ $\mu$s; at later times, the pair correlations do not show any macroscale structure within the displayed field of view. The center regions (distances $\lesssim$ 100 $\mu$m) of each pair correlation reveal the short-range correlations between ions with their nearest neighbors, as seen by the lighter region surrounding the center. These correlations are due to Coulomb repulsion. The range within which ion pairs are under-abundant (correlation $<$ 1) increases over time, meaning that the minimum separation between repelling ion pairs increases with time. These correlations are a hallmark of strong coupling and are closely related with disorder-induced heating \cite{chen04}, which is observed in the present work by spatially-resolved ion imaging.

Furthermore, inspection of Fig.~\ref{fig:pics_low} shows that the ion separation increases faster in the vertical than it does in the horizontal direction. This is due to the overall geometry of our plasma. The axis of the plasma cylinder is aligned horizontally across the photos, and the comparatively smaller radius of the cylinder stretches out vertically. As ions pairs separate from each other axially, they eventually interact with initially distant ions that suppress further expansion. Conversely, in the radial directions the ions expand into free space, unimpeded by other ions.

\begin{figure}
    \centering
    \subfloat[]{
        \includegraphics[width=4 cm]{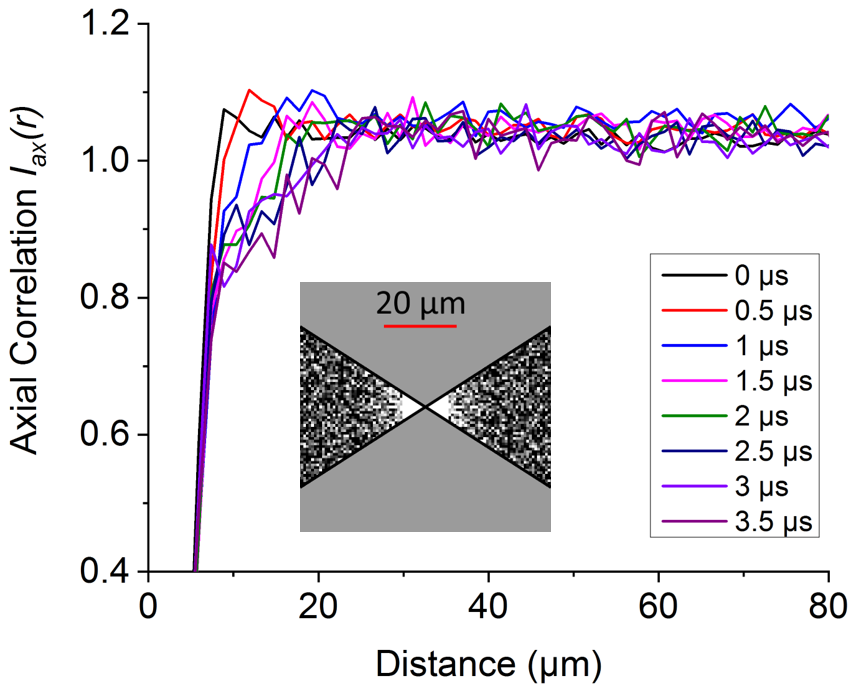}
        \label{subfig:integrals_ax_exp}
        }
    \subfloat[]{
        \includegraphics[width=4 cm]{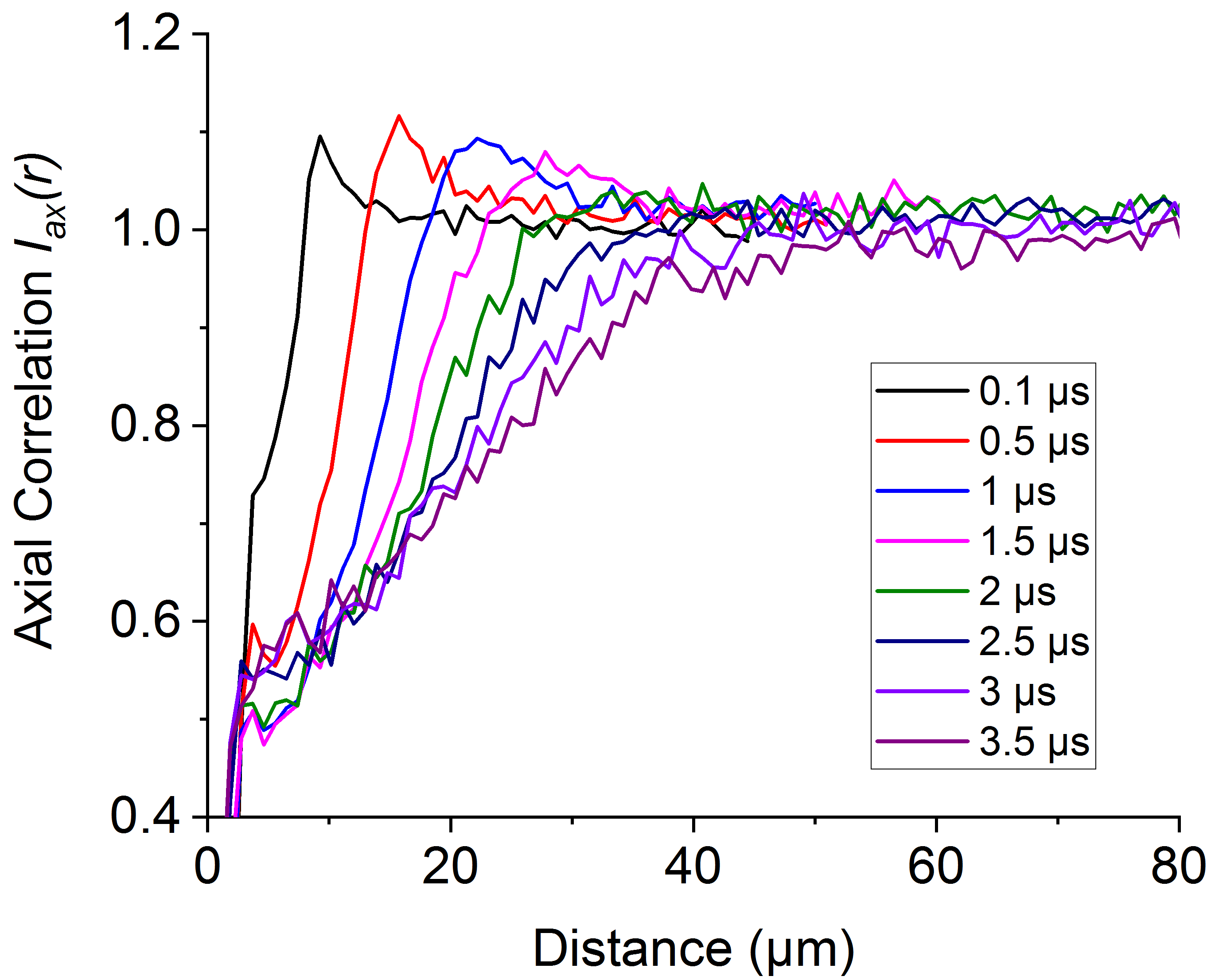}
        \label{subfig:integrals_ax_comp}
        }
    \linebreak
    \subfloat[]{
        \includegraphics[width=4 cm]{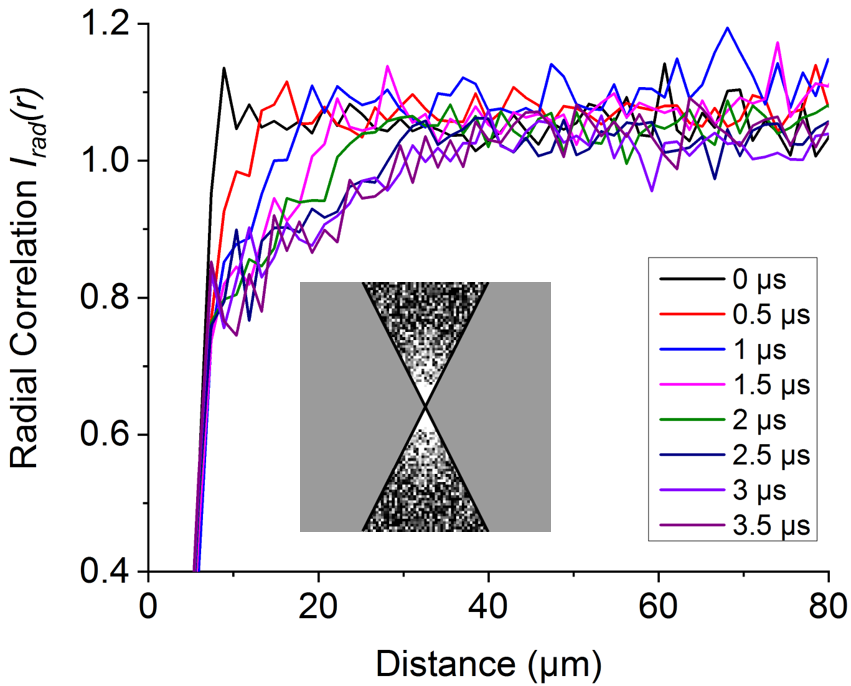}
        \label{subfig:integrals_rad_exp}
        }
    \subfloat[]{
        \includegraphics[width=4 cm]{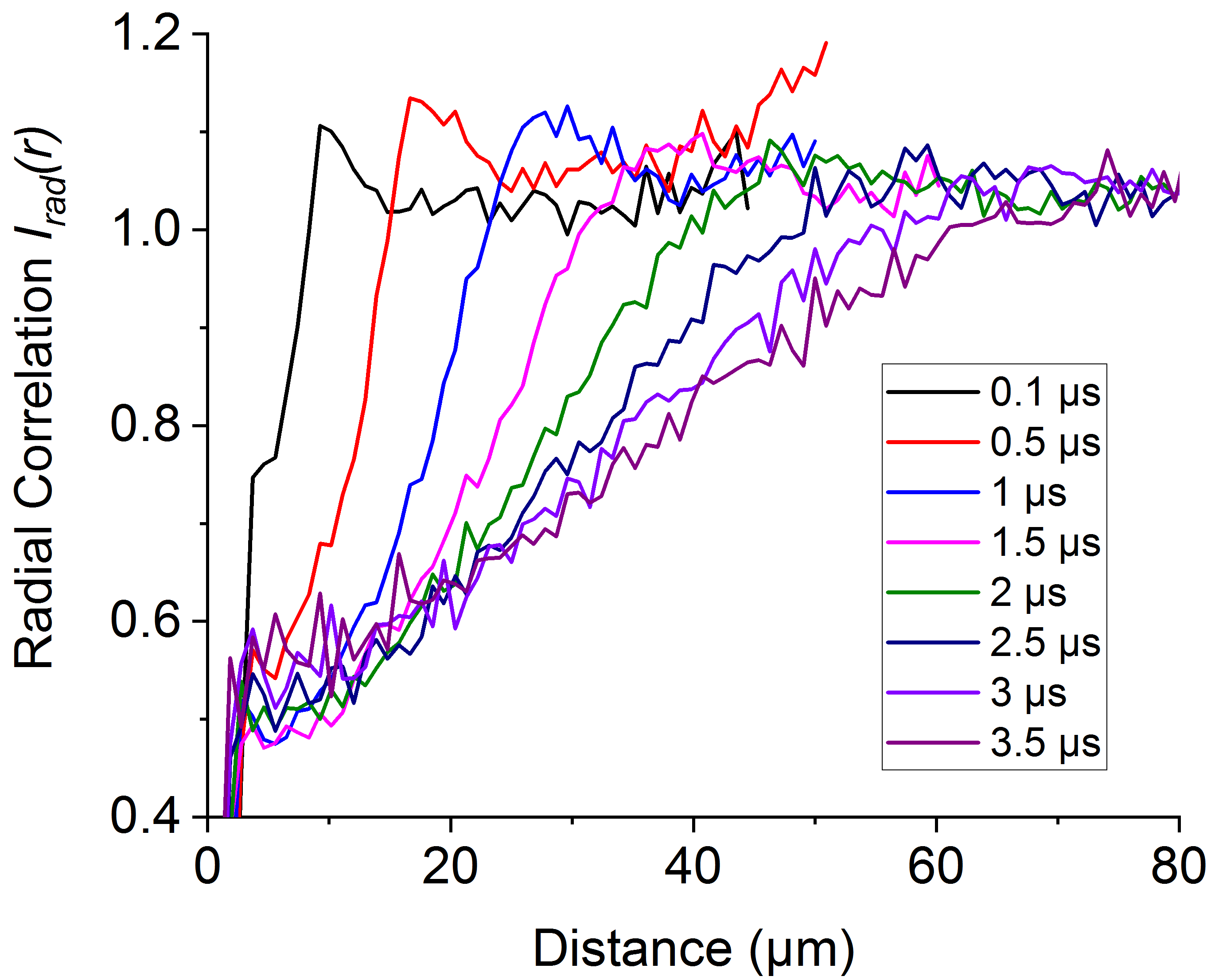}
        \label{subfig:integrals_rad_comp}
        }
    \caption{Experimental (\ref{subfig:integrals_ax_exp} and \ref{subfig:integrals_rad_exp}) and computational (\ref{subfig:integrals_ax_comp} and \ref{subfig:integrals_rad_comp}) axial and radial angular integrals of the pair correlation functions for each time step. The inset images are the integration regions. Each integral was performed over the specified region to show the difference in pair correlations in the radial and axial directions.}
    \label{fig:integrals}
\end{figure}

To highlight the difference between radial and axial correlations, we produce angular integrals of the pair correlations. Figure~\ref{fig:integrals} shows experimental and computational angular integrals, $I_{\textrm{ax}}(r)$ and $I_{\textrm{rad}}(r)$, along with the corresponding integration regions. Experimental and corresponding simulated results agree very well in terms of the triangular shape of the pair-correlation functions, their gradual expansion to larger distances, and the presence of enhanced correlation at $\tau \lesssim 1$ $\mu$s in a range 10 $\mu$m $\lesssim r \lesssim$ 20 $\mu$m, where $I(r) > 1$. Just as the pair correlation images suggest, we see that the regions of anti-correlation ($I(r) < 1$) expand faster in the radial than the axial direction. To further quantify this, we apply two linear fits to each trace as illustrated in the inset image in Table~\ref{tab:separations}. We then calculate the intersection point of the two linear fits to find the correlation lengths, $\rho_{\textrm{ax}}$ and $\rho_{\textrm{rad}}$, for the axial and radial directions respectively. These lengths are listed in Table~\ref{tab:separations}, along with their ratios. 

We note that the simulated correlation lengths, $\rho_{\textrm{axc}}$ and $\rho_{\textrm{radc}}$, are greater than the measured correlation lengths by up to 30\% before the last time step. We owe this discrepancy to image distortion due to higher-order multipole field effects in the ion-imaging system. We also see that the measured and computed correlation functions $I(r)$ drop below 0.4 as $r \rightarrow 0$. These drop-offs are artifacts that occur at different distances. In the experimental plots, the drop-off occurs at $\approx 6~\mu$m because of the ion blip size on the MCP (in the object plane). In the simulated plots, the drop-off occurs at $\approx 2~\mu$m, which is the bin size in the simulation.

\begin{table}
    \vspace{0 pt}
    \centering
    \includegraphics[width=6 cm]{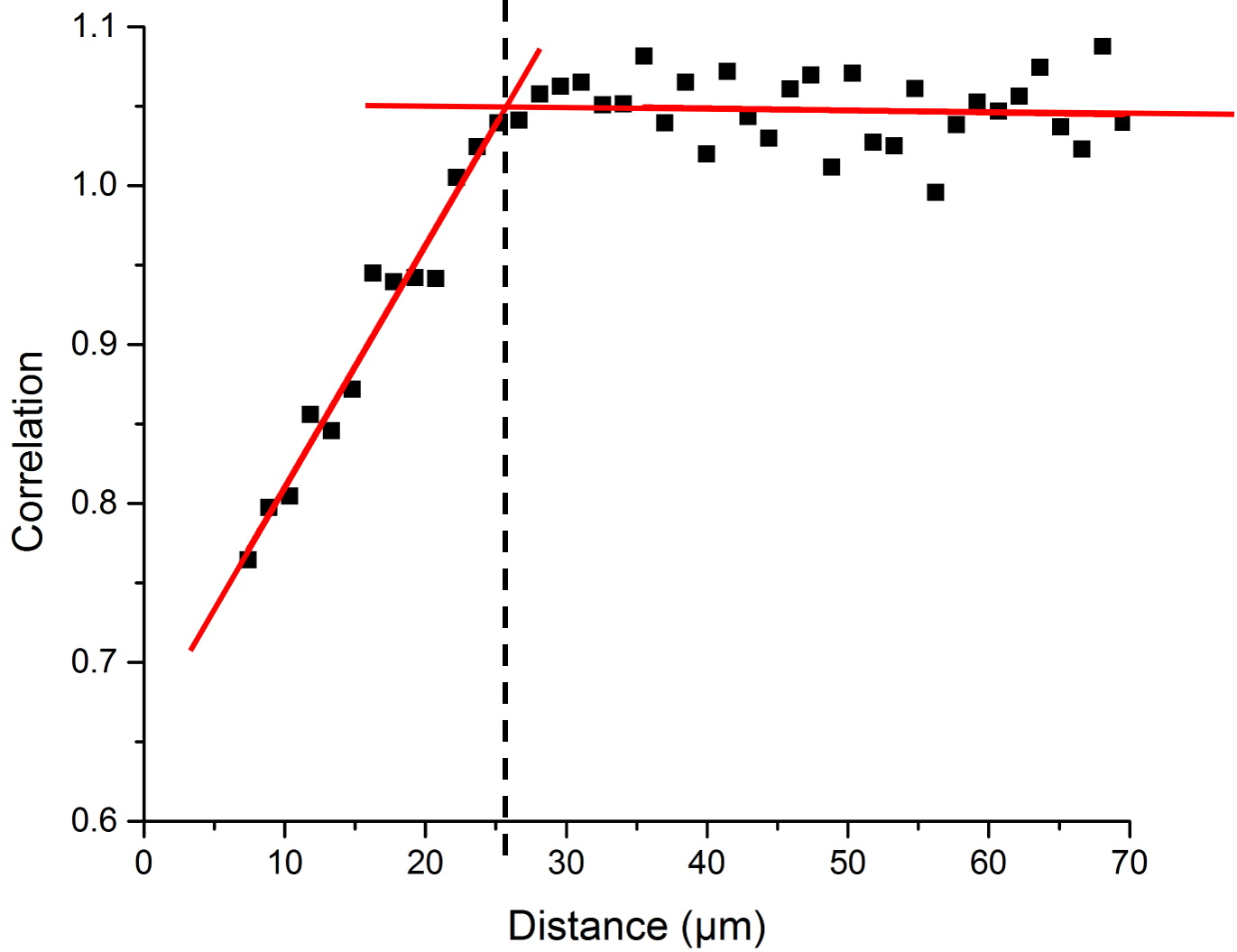}
    \vspace{0 pt}
    \footnotesize
    \begin{tabular}{c||c|c||c|c||c|c}
        \makecell{Expansion \\ Time [$\mu$s]} & \multicolumn{2}{c||}{Axial [$\mu$m]} & \multicolumn{2}{c||}{Radial [$\mu$m]} & \multicolumn{2}{c}{\makecell{Radial/Axial \\ Ratio}} \\
        {} & $\rho_{\textrm{ax}}$ & $\rho_{\textrm{axc}}$ & $\rho_{\textrm{rad}}$ & $\rho_{\textrm{radc}}$ & Exp. & Comp. \\
        \hline
        0.0 & 7.70 & -- & 8.37 & -- & 1.09 & -- \\
        0.1 & -- & 8.66 & -- & 8.62 & -- & 1.00 \\
        0.5 & 10.10 & 13.68 & 13.95 & 15.19 & 1.38 & 1.11 \\
        1.0 & 14.98 & 18.31 & 19.76 & 24.13 & 1.32 & 1.32 \\
        1.5 & 16.37 & 21.74 & 25.21 & 32.11 & 1.54 & 1.48 \\
        2.0 & 19.70 & 25.68 & 28.15 & 38.73 & 1.43 & 1.51 \\
        2.5 & 23.09 & 30.02 & 35.53 & 50.64 & 1.54 & 1.69 \\
        3.0 & 25.36 & 35.28 & 38.16 & 59.27 & 1.50 & 1.68 \\
        3.5 & 26.76 & 38.71 & 42.76 & 67.17 & 1.60 & 1.74 
    \end{tabular}
    \caption{Measured axial and radial correlation lengths, $\rho_{\textrm{ax}}$ and $\rho_{\textrm{rad}}$, vs. time, and respective simulated values, $\rho_{\textrm{axc}}$ and $\rho_{\textrm{radc}}$, obtained from the angular integrals in Fig.~\ref{fig:integrals}. The inset on top shows how the correlation lengths are obtained. The last two columns show the measured and simulated ratios between the correlation lengths.}
    \label{tab:separations}
\end{table}

\section{Conclusion} \label{sec:conclusion}

We have observed strong-coupling effects of an ion plasma that adiabatically expands into the vacuum, and compared our results with two models. The measured indicators for strong couplings are the formation of shock fronts, seen in measutred and calculated time-of-flight signals, and short-range correlations between nearest and next-nearest neighbors, evidenced by non-symmetric pair correlation functions with regions of reduced and enhanced correlation. Strong coupling appears to persist throughout the investigated expansion time scale, because the measured indicators for strong couplings persist. However, it is expected that disorder-induced heating reduces the coupling parameter from its initial value $\Gamma \sim 30$ to a lower value (that is still larger than one.) Our trajectory simulations indicate a velocity increase from about 2~m/s to about 6~m/s over the first few $\mu$s, which may translate into a drop of $\Gamma$ to near three.

We have used different density regimes of the plasma to exhibit different strong-coupling effects. Higher-density plasma behave more fluid-like, and the ion arrival times reveal shock fronts and chamber-specific focusing effects. Lower-density plasmas, meanwhile, reveal directionally-dependent pair correlations; these are likely also present but obscured in larger plasmas (due to image projection onto a plane). In future work, we plan on observing non-neutral plasmas with different initial conditions, investigating neutral plasmas, and applying atom-based electric-field metrology techniques to cold plasmas. This may enable a distinction between microscopic (Holtsmark) and macroscopic fields. To push the microscopic fields to higher values, amenable to experimental detection, one may use the plasma focus effect described in Fig.~\ref{fig:quadupole} to generate a transient spike in plasma density.

\begin{acknowledgments}
This work was supported by NSF Grant PHY-1707377. We thank David Anderson, Rydberg Technologies Inc, and Eric Paradis, Eastern Michigan University, for valuable discussions.
\end{acknowledgments}


\bibliography{plasmaexp}

\end{document}